\documentclass[twocolumn]{aastex6}

\usepackage{import}
\usepackage{color}
\usepackage{listings}
\usepackage{amsmath}
\DeclareMathOperator{\taninv}{arctan}

\begin{document}

\title{\orbitize{}: A Comprehensive Orbit-fitting Software Package for the High-contrast Imaging Community}
\author{
    Sarah Blunt\altaffilmark{1,2}, 
    Jason Wang\altaffilmark{1,3}, 
    Isabel Angelo\altaffilmark{4}, 
    Henry Ngo\altaffilmark{5}, 
    Devin Cody\altaffilmark{6},
    Robert J. De Rosa\altaffilmark{7},
    James Graham\altaffilmark{8},
    Lea Hirsch\altaffilmark{7},
    Vighnesh Nagpal\altaffilmark{9},
    Eric L. Nielsen\altaffilmark{7},
    Logan Pearce\altaffilmark{10,2},
    Malena Rice\altaffilmark{11,2},
    Roberto Tejada\altaffilmark{12,7}
}

\altaffiltext{1}{Department of Astronomy, California Institute of Technology, Pasadena, CA, USA}
\altaffiltext{2}{NSF Graduate Research Fellow}
\altaffiltext{3}{51 Pegasi b Fellow}
\altaffiltext{4}{Department of Physics and Astronomy, University of California, Los Angeles, Los Angeles, CA, USA}
\altaffiltext{5}{NRC Herzberg Astronomy \& Astrophysics Research Centre, Victoria, BC, Canada}
\altaffiltext{6}{Jet Propulsion Laboratory, California Institute of Technology, Pasadena, CA}
\altaffiltext{7}{Kavli Institute for Particle Astrophysics and Cosmology, Stanford University, Stanford, CA, USA}
\altaffiltext{8}{Department of Astronomy, University of California, Berkeley, Berkeley, CA, USA}
\altaffiltext{9}{Lexington High School, Lexington, MA, USA}
\altaffiltext{10}{Steward Observatory, University of Arizona, Tucson, AZ, USA}
\altaffiltext{11}{Department of Astronomy, Yale University, New Haven, CT, USA}
\altaffiltext{12}{California State University, Los Angeles, Los Angeles, CA, USA}

\shortauthors{Blunt et al.}
\shorttitle{orbitize!}
\submitted{Submitted to AAS Journals}

\newcommand{\mstar}{\ensuremath{M_\star}}
\newcommand{\msun}{\ensuremath{M_\odot}}
\newcommand{\mjup}{\ensuremath{M_J}}
\newcommand{\ms}{\ensuremath{\mathrm{ms^{-1}}}}
\newcommand{\orbitize}{\texttt{orbitize!}}
% make red text with \red{some text} - note: doesn't work with paragraph breaks
\newcommand{\red}[1]{\textcolor{red}{#1}}

\newcommand{\currentversion}{1.6.0}

\begin{abstract}
    \orbitize{} is an open-source, object-oriented software package for fitting the orbits of directly-imaged objects. It packages the Orbits for the Impatient (OFTI) algorithm and a parallel-tempered Markov Chain Monte Carlo (MCMC) algorithm into a consistent and intuitive Python API. \orbitize{} makes it easy to run standard astrometric orbit fits; in less than 10 lines of code, users can read in data, perform one fit using OFTI and another using MCMC, and make two publication-ready figures. Extensive pedagogical tutorials, intended to be navigable by both orbit-fitting novices and seasoned experts, are available on our documentation page. We have designed the \orbitize{} API to be flexible and easy to use/modify for unique cases. \orbitize{} was designed by members of the exoplanet imaging community to be a central repository for algorithms, techniques, and know-how developed by this community. We intend for it to continue to expand and change as the field progresses and new techniques are developed, and call for community involvement in this process. Complete and up-to-date documentation is available at \href{http://orbitize.info/en/latest/}{orbitize.info}.
\end{abstract}

\keywords{open source software - exoplanet detection methods: coronographic imaging - orbit determination}

\section{Introduction}
\label{sec:intro}

By repeatedly imaging exoplanets, we can directly observe them moving along their orbits. The physics behind orbits is well-established and straightforward to computationally model, and can reveal much about the properties and dynamical histories of planetary systems. For example, misalignment of the orbital plane with the stellar spin axis can indicate disturbances early in the lifetime of a system before planets formed \citep{Bate:2010aa, Maire:2019aa}. In systems with circumstellar dust, a planet can warp a disk when the two are mutually inclined \citep{Lagrange:2010aa,Dawson:2011aa} or scatter comets into the inner planetary system \citep{Thebault:2001aa,Zieba:2019aa}. Planet-planet scattering and resonant migration in a gas disk can excite observable eccentricities \citep{Yu:2001aa,Scharf:2009aa}, potentially implying the presence of unseen planets in planetary systems and constraining formation scenarios. Improved astrometric precision \citep[e.g.,][]{Gravity-Collaboration:2019aa} could soon lead to the detection of unseen planets based on perturbations in the orbits we observe. In the future, accurate orbit models of exo-Earths from future space imaging missions will be critical to properly assess their climates and habitability \citep{Williams:2002aa}.

While the physics of orbits is straightforward, orbit-fitting is challenging, especially for the current generation of directly imaged planets. The wide angular separations needed to detect these objects translate to decades-long or longer orbital periods, which means that the parameter space of possible orbits is often large. Orbit-fitting packages such as \texttt{ExoFast} \citep{Eastman:2013aa} and PyAstrOFit \citep{Wertz:2017aa} explore these parameter spaces using a Markov-Chain Monte Carlo (MCMC) approach, but such algorithms often converge slowly in this regime. Several specialized techniques have been developed in the past few years to address some of these difficulties (e.g. \citealt{Blunt:2017aa}, \citealt{ONeil:2019aa}, \citealt{Brandt:2018aa}), but it is often left as an exercise to the reader to implement, debug, and combine these techniques. 

In this paper, we present \texttt{orbitize!}, an open-source orbit-fitting software package inspired by \texttt{radvel} \citep{Fulton:2018aa}, designed to meet the needs of the high-contrast imaging community. \texttt{orbitize!} is designed to consolidate the algorithms, techniques, and know-how of the high-contrast orbit-fitting community in one place. It is fast and robust, but also clearly written, well-documented, and easy to use. We have designed \texttt{orbitize!} to be flexible and easily modifiable so it can grow with the field of high-contrast orbit-fitting. We believe \texttt{orbitize!} to be a code base with a comparatively low barrier to understanding and contribution. We seek to remove obstacles to becoming an expert in direct-imaging orbit-fitting, enabling the field to advance more rapidly. 

This paper is organized as follows: in Section \ref{sec:orbitfitting}, we review Bayesian orbit-fitting for directly imaged astrometry and discuss the various algorithms used to perform this procedure. In Section \ref{sec:code}, we outline the design of \texttt{orbitize!} and give examples of how to use the code. In Section \ref{sec:future}, we discuss community involvement guidelines and provide a list of items on the \texttt{orbitize!} to-do list, and we conclude in Section \ref{sec:concl}.

\section{Fitting Imaged Orbits}
\label{sec:orbitfitting}

\subsection{Defining the Orbit}
\label{sec:orbitelements}

\begin{figure}
    \centerline{\includegraphics[width=0.5\textwidth]{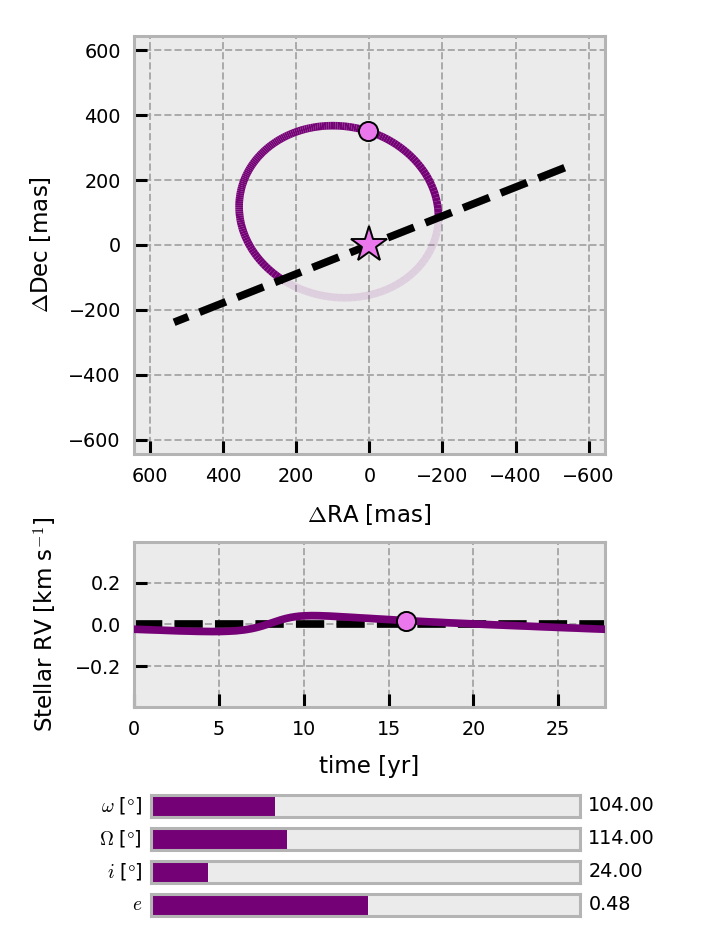}}
    \caption{Screen-capture of an animated, interactive Python module intended to help users visualize orbits and the \orbitize{} coordinate system. Top: RA vs decl. The orbit's line of nodes is shown as a dashed line, and the primary's location as a pink star. The dark purple arc shows the portion of the orbit in front of the sky plane, and the lighter purple arc shows the portion that is behind the sky plane. Middle: stellar radial velocity versus time. Bottom: interactive sliders that users can drag to set the values of orbital elements in the interactive version. This figure is available as a video in the online version of this article. In addition, readers are encouraged to download and interact with this visualization at \href{https://github.com/sblunt/orbitize/blob/master/docs/tutorials/show-me-the-orbit.ipynb}{this link}.}
    \label{fig:angles-gif}
\end{figure}

Orbits in \texttt{orbitize!} are parametrized using the angle conventions from \citet{Green:1985aa}: semi-major axis ($a$), eccentricity ($e$), inclination angle ($i$), argument of periastron of the secondary's orbit ($\omega$), longitude of ascending node ($\Omega$), epoch of periastron passage ($\tau$), parallax, and total mass. Note that we express epoch of periastron passage ($\tau$) as a fraction of the orbital period past a specified reference date $t_{\mathrm{ref}}$ (default January 1, 2020):

\begin{equation}
    \tau = \left( \frac{t_0 - t_{\mathrm{ref}}}{P} \right) ~\textrm{mod}~1,
\end{equation}
where $t_0$ is the time of periastron and $P$ is the orbital period. We chose to fit in $\tau$ rather than $t_0$ because the prior bounds for $\tau$ are straightforward (between 0 and 1) no matter the orbital period. A screen-capture from an interactive module intended to help readers visualize these parameters is shown in Figure \ref{fig:angles-gif}, and an interactive version is available online\footnote{\href{https://github.com/sblunt/orbitize/blob/master/docs/tutorials/show-me-the-orbit.ipynb}{github.com/sblunt/orbitize/blob/master/docs/tutorials/show-me-the-orbit.ipynb}}.

In the \orbitize{} coordinate system, motion along the positive z-direction causes a redshift. The positive x-direction is in the positive R.A. direction offset from the primary star, and the positive y-direction is in the positive decl. direction. The orbital elements are defined as usual within this reference frame, with i=0$^{\circ}$ corresponding to a face-on orbit. We caution, however, that users must be careful comparing the outputs of RV orbit-solving packages like \texttt{radvel} with those of \orbitize{}, since \texttt{radvel} fits the orbit of the star, and \orbitize{} fits the orbit of the planet. In practice, this just means adding $\pi$ to any argument of periastron ($\omega$) values returned by an RV code. See Figure 1 of \citet{Fulton:2018aa} for a visualization of this difference. 

\begin{deluxetable}{lccc}
\centering
\tabletypesize{\scriptsize}
\tablewidth{0pc}
\tablenum{1}
\tablecaption{Orbit Model Parameters}
\tablehead{\colhead{Parameter} & \colhead{Symbol} & \colhead{Unit} & 
\colhead{Default Prior}}
\startdata
Semi-major axis & $a$ & au & Log-Uniform\\
Eccentricity & $e$ & & Uniform[0,1] \\ 
Inclination & $i$ & rad & Sine[0,$\pi$]\\
Argument of Periastron& $\omega$ & rad & Uniform[0,2$\pi$]\\
Longitude of Ascending Node& $\Omega$ & rad & Uniform[0, 2$\pi$]\\
Epoch of Periastron Passage$^a$ & $\tau$& & Uniform [0,1]
\enddata
\tablenotetext{a}{Expressed as fraction of orbital period past Jan 1, 2020}
\label{tab:orbital_parameters}
\end{deluxetable}

\subsection{Bayesian Orbit Fitting}
\label{sec:bayesian_estimation}

For high-contrast imaging, Bayesian orbit-fitting is the process of converting time series measurements of a planet's location relative to its primary to a posterior over its orbital parameters. The inputs to this process are times at which measurements were taken and relative position measurements, most often expressed as a planet-star separation ($\rho$) and position angle ($\theta$), or as offsets in R.A. ($\alpha$) and decl. ($\delta$).

To compute the posterior over orbital parameters, we use Bayes' theorem:

\begin{equation}
    p(m | \mathcal{D}) \propto \mathcal{L}(m | \mathcal{D})p(m)
\end{equation} where $p(m | \mathcal{D})$ is the posterior, $\mathcal{L}(m | \mathcal{D})$ is the likelihood (the probability of the orbit model given the data), and $p(m)$ is the prior (the probability of the orbit model itself). By default, \texttt{orbitize!} uses a Gaussian likelihood:

\begin{equation}
    \log{\mathcal{L}(m | \mathcal{D})} = -\frac{1}{2}\chi^2_{\theta} -\frac{1}{2}\chi^2_{\rho}
    -\frac{1}{2}\chi^2_{\alpha}
    -\frac{1}{2}\chi^2_{\delta}
\end{equation}

\begin{equation}
\label{eq:theta}
    \chi^2_{\theta} = \sum_{i=1}^{N} \frac{\left[\taninv\left(\frac{\sin(\theta_m(t_i)-\theta_o(t_i))}{\cos(\theta_m(t_i)-\theta_o(t_i))}\right)\right]^2}{ \sigma_{\theta_o(t_i)}^2}
\end{equation}

\begin{equation}
    \chi^2_{\rho} = \sum_{i=1}^{N} \frac{(\rho_m(t_i) - \rho_o(t_i))^2}{ \sigma_{\rho_o(t_i)}^2}
\end{equation}

\begin{equation}
    \chi^2_{\alpha} = \sum_{i=1}^{M} \frac{(\alpha_m(t_i) - \alpha_o(t_i))^2}{ \sigma_{\alpha_o(t_i)}^2}
\end{equation}

\begin{equation}
    \chi^2_{\delta} = \sum_{i=1}^{M} \frac{(\delta_m(t_i) - \delta_o(t_i))^2}{ \sigma_{\delta_o(t_i)}^2}
\end{equation} where $N$ is the number of observation epochs with observations measured in terms of $\rho$ and $\theta$, $M$ is the number of observation epochs with observations measured in terms of $\alpha$ and $\delta$, $t_i$ is the epoch of the $i$th observation, $\rho_m(t_i)$ is the separation predicted by the model orbit at $t_i$, $\rho_o(t_i)$ is the observed separation at $t_i$, and $\sigma_{\rho_o(t_i)}$ is the observational uncertainty on the observed separation. The inverse tangent function in Equation \ref{eq:theta} accounts for angle wrapping near 0 and $2\pi$ to ensure that the difference between the model and observed value is calculated correctly. $\rho_m(t_i)$, $\theta_m(t_i)$, $\alpha_m(t_i)$, and $\delta_m(t_i)$ are determined by solving Kepler's equation, discussed further in Section \ref{sec:kepler}. 

Total mass and parallax are either included as free parameters and assigned priors motivated by observations or held fixed in this analysis. In \orbitize{}, users can select either option.

\subsection{Algorithms for Orbit-fitting}

Prior to more modern Bayesian techniques, well-constrained visual orbits were typically fit with least-square minimization techniques (e.g. \citealt{Binnendijk:1960aa}, \citealt{Sozzetti:1998aa}). While such methods are computationally efficient, and they effectively find the maximum likelihood orbit with estimates of the 1D errors and 2D covariances, they do not produce full posteriors. This is particularly problematic for poorly constrained orbits, when the posteriors are distinctly non-Gaussian.

A more computationally intensive method is Least-Square Monte Carlo (LSMC), which generates input orbits by drawing randomly from a multi-dimensional parameter space of orbital parameters. Each of the input orbits is run through an iterative $\chi^2$ minimizer until a stopping criterion is reached \citep[e.g. ][]{Mugrauer:2012aa}. Several recent publications \citep{Ginski:2013aa, Vigan:2016aa, Maire:2019aa} have demonstrated that that while LSMC is capable of finding families of plausible orbits, its output is often significantly different from the Bayesian posterior. Another Monte Carlo based method was used in \citet{Wagner:2016aa} and \citet{Wagner:2018aa} to fit short orbital arcs. This method used a multi-step grid-search algorithm to find the maximum likelihood orbit, and Gaussian point estimates to approximate confidence intervals. 

Both LSMC and grid search algorithms have difficulty fitting systems with data covering only short orbital arcs, in part because both prioritize finding the set of orbital parameters corresponding to the maximum likelihood solution. In a well-constrained orbit where the posterior is a six-dimensional Gaussian, either method should find the best-fitting parameters and recover the marginalized uncertainty in each parameter. When the posteriors are decidedly non-Gaussian, however (which is much more often the case for orbit fitting in direct imaging), both techniques fail to adequately derive the shape of the posterior. Additionally, neither of these methods appears to offer a gain in computational efficiency compared to Orbits for the Impatient (OFTI; see below) or MCMC.

% For example, \citet{Maire:2019aa} fit the orbit of 51 Eridani b using LSMC with 1,000,000 initial orbits. For a Levenberg-Marquardt solver, the number of times for which separation and position angle would need to be evaluated for each initial orbit is the number of iterations of the solver times the number of parameters times 2 (to calculate the numerical partial derivatives). Thus, the number of orbit calculations is likely to be of order 100 per initial orbit, or $10^8$ total orbit calculations. As another example, \citet{Wagner:2018aa} described a method that required calculating $46,656 \times 7 \times 25,265 = 8.3\times10^{9}$ sets of orbital parameters, of which 25,265 were kept. For comparison, for a typical case where the OFTI acceptance rate is 10\%, orbital parameters would be calculated $10^7$ times to generate a set of $10^6$ orbital parameters.

In the following two subsections, we describe OFTI and MCMC, the two backend algorithms available in \orbitize{}. These algorithms represent efficient Bayesian methods for producing plausible sets of orbital parameters that represent the full multi-dimensional posterior of orbital parameters. These posteriors are robust probability density functions with confidence intervals that are an accurate reflection of the constraints imposed by the data.

\subsubsection{Orbits for the Impatient (OFTI)}
\label{sec:OFTI_algorithm}

The OFTI algorithm is described in detail in \citet{Blunt:2017aa}, but we review it briefly here. Trial orbits are drawn randomly from priors, and are ``scaled-and-rotated'' to match the data. Scaling-and-rotating involves modifying the semimajor axis and position angle of nodes of the trial orbit to match the most constrained astrometric data point within the observational uncertainties, cutting down the large parameter space of possible orbits. For each scaled-and-rotated trial orbit, Kepler's equation is solved (See Section \ref{sec:kepler}) and a likelihood is computed. Finally, each orbit is either accepted or rejected by comparing the likelihood probability to a uniform random number. 

The OFTI algorithm is most efficient when the orbital posteriors are similar to the priors, or in other words when the parameter space is relatively unconstrained. This occurs most often for short orbital arcs, when the data span only a small fraction of the total orbital period. 

In the case of OFTI, each individual orbit considered is uncorrelated with the rest, and so the only stopping criterion is the number of samples desired, which varies by application. To plot plausible orbit tracks, $\sim$100 sets of orbital parameters are sufficient. $\sim$1000 may be needed for accurate medians and 1$\sigma$ confidence intervals on each parameter. Plots of 1D marginalized posteriors are well-sampled with $\sim 10^4$ sets, and 2D parameters with $\sim 10^6$.

\subsubsection{Affine-Invariant \& Parallel Tempered MCMC}
\label{sec:MCMC_algorithm}
Markov-chain Monte Carlo (MCMC) algorithms are commonly used to sample the posterior of planetary orbits. \orbitize{} makes use of two such algorithms: the Affine-invariant sampler from \texttt{emcee} \citep{Foreman-Mackey:2013aa}, and the parallel-tempered MCMC (PTMCMC) sampler from \texttt{ptemcee} \citep{Vousden:2016aa}. Given the complex covariances and often multi-modal posteriors of the orbits of directly-imaged planets, the Affine-invariant sampler alone generally fails to fully sample the posterior without fine-tuned starting locations for the walkers. We offer the use of \texttt{ptemcee}, which runs multiple Affine-invariant samplers with different likelihood weights, as an alternative to overcome this difficulty.

MCMC algorithms generally have similar run times for orbits with partial phase coverage \citep{Blunt:2017aa}. However, convergence time is cut down significantly if the orbital elements are well-constrained, such that the posterior is close to a multivariate Gaussian. 

Unlike OFTI, the MCMC algorithms coded in \orbitize{} require an initial period for the walkers to fully converge before they sample the posterior in an unbiased fashion. Convergence is assessed using a combination of the walkers' autocorrelation time as recommended by \citet{Foreman-Mackey:2013aa} and visual inspection of the walker positions over time to determine if they are fully exploring parameter space. Unconverged posteriors typically appear ``lumpy,'' as different chains are still in different regions of parameter space, compared to generally smooth (though not necessarily Gaussian) posteriors for converged chains. See Figure 4 of \citet{Blunt:2017aa} for an illustration. The multimodal posteriors of $\omega$ and $\Omega$ can also illustrate convergence. Without a measurement of radial position or velocity, values $180^{\circ}$ apart in both $\omega$ and $\Omega$ are degenerate. Thus, the chains are more likely to have converged when the one-dimensional marginalized posteriors on $\omega$ and $\Omega$ show symmetric equal peaks. 

\section{Code Design}
\label{sec:code}

\subsection{API}
\label{sec:API}

\texttt{orbitize!} comprises several modules, each a separate code file. This section refers to \texttt{orbitize!} version \currentversion, but the most up-to-date documentation can be found online\footnote{\href{http://orbitize.info/en/latest/}{orbitize.info}}. The primary modules are:

\subsubsection{Input Data}
The \texttt{orbitize.read\_input} module is designed to read astrometric measurements as input to \orbitize\ in any file format supported by \texttt{astropy.io.ascii.read()}, including csv, fixed width, cds, and LaTeX formats. The main method in this module is \texttt{orbitize.read\_input.read\_file()}.

This module also contains the method \texttt{orbitize.read\_input.write\_orbitize\_input()} which takes a table of measurements formatted in the \orbitize\ format and writes it as an ASCII file in any file format supported by \texttt{astropy.io.ascii.read()}. 

% Finally, there are two deprecated methods in this module. \texttt{orbitize.read\_input.read\_formatted\_file} and \texttt{orbitize.read\_input.read\_orbitize\_input} formerly provided different methods to read different input types. Now, these functions are just wrappers to \texttt{orbitize.read\_input.read\_file} which can handle both formats. These functions will be removed in the v2.0 release of \orbitize.

\subsubsection{Priors}
Priors in \orbitize{} are represented as subclasses of the abstract super class \texttt{orbitize.priors.Prior}. Each prior class can draw random values from its distribution as well as compute the probability of drawing particular values from that prior distribution. Other parts of the code use these two methods to interface with the priors. Currently, \orbitize{} implements uniform, Gaussian, log-uniform (Jeffreys), sine, and linear priors. The default priors listed in Table \ref{tab:orbital_parameters} are used for orbit fits, but users can easily replace priors by instantiating their own priors and replacing them in the \texttt{orbitize.system.System} instance.\footnote{\href{http://orbitize.info/en/latest/tutorials/Modifying\_Priors.html}{http://orbitize.info/en/latest/tutorials/Modifying\_Priors.html}}

\subsubsection{Kepler's Equation Solver}
\label{sec:kepler}
The \texttt{orbitize.kepler} module converts orbital elements to position and radial velocity measurements by generating an orbital ellipse, placing the companion at the appropriate phase of the orbit, and rotating the orbit based on viewing geometry. The instantaneous position of a companion along an orbital ellipse is defined by the true anomaly $\nu$, the angle between the location of the companion and periastron, as measured from the focus of the orbital ellipse. $\nu$ cannot be analytically calculated from a given epoch and set of orbital elements. It is instead calculated from the eccentric anomaly $E$, the angle between the projection of the companion's location onto a circle of radius $a$ that intersects the true orbit at periastron, which is calculated in turn from the mean anomaly $M$, the fraction of the period that has elapsed since periastron passage. See \citet{Seager:2010aa} for an excellent derivation of the relevant equations from first principles and visualizations of these quantities.

\begin{figure*}
    \centerline{\includegraphics[width=\textwidth]{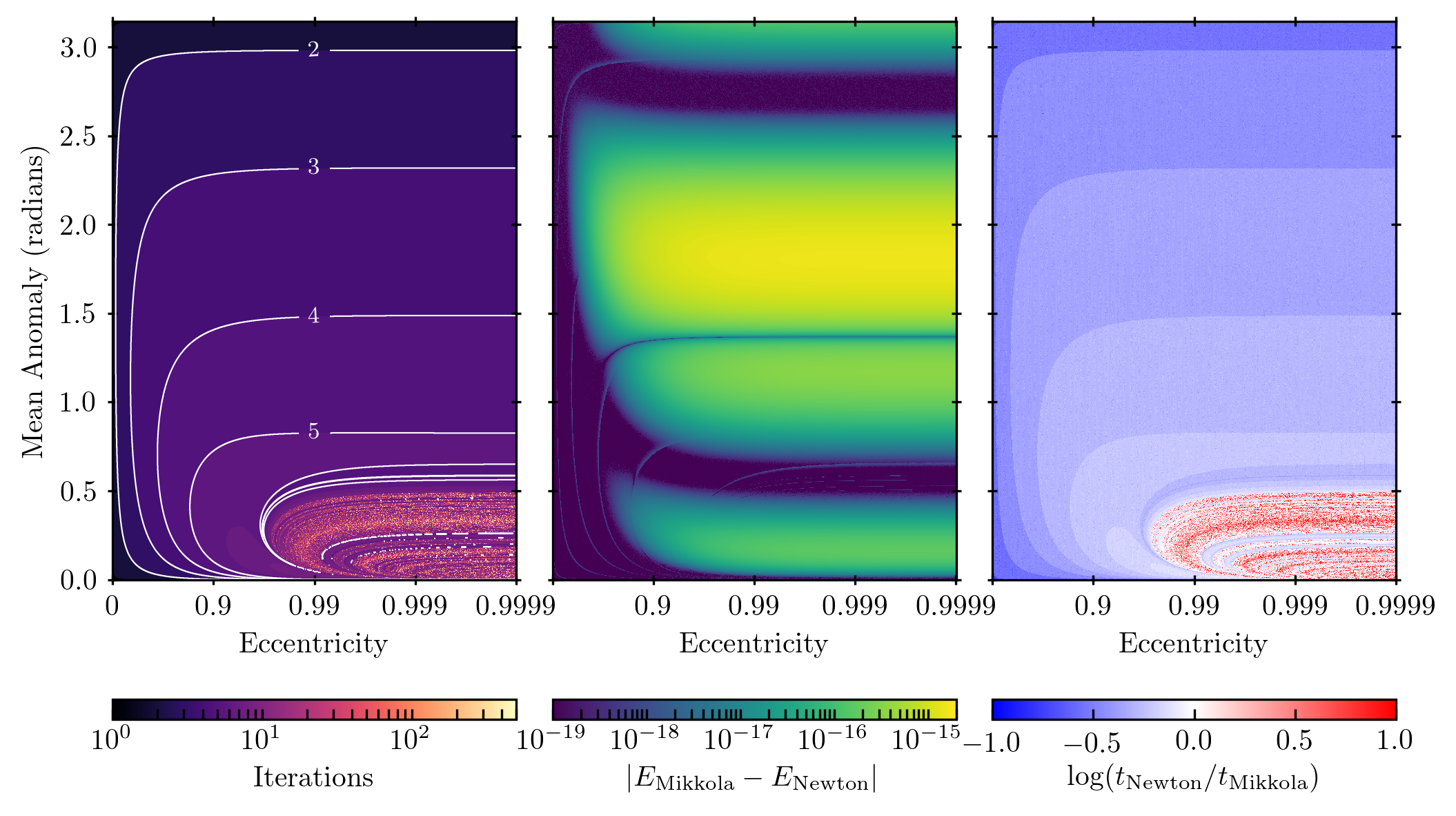}}
    \caption{Performance summary of our two Keplerian solvers. Left: Number of iterations by the Newton solver required to achieve a tolerance of 1e-9 as a function of eccentricity and mean anomaly. The Mikkola solver is used for orbits with $e\ge0.95$ to mitigate this slowdown at high eccentricities. Middle: absolute difference between the eccentric anomaly values computed by the Mikkola and Newton solvers. A tolerance of 1e-18 was used in this experiment. The absolute difference between the two solvers is negligible compared with current astrometric precision. Right: time needed for both solvers to achieve a tolerance of 1e-18. The Mikkola solver achieves significant performance gains over the Newton solver for high eccentricities.}
    \label{fig:kepler}
\end{figure*}

The conversion between $M$ and $E$, typically called ``Kepler's equation,'' is given by:

\begin{equation}
    M = E - e\sin{E}.
\end{equation} This equation cannot be solved analytically. In \orbitize{}, we solve Kepler's equation using one of two different methods depending on the eccentricity of the orbit. For $e<0.95$, we use Newton's method to estimate $E$ with a default tolerance of $|E_n - E_{n-1}| < 10^{-9}$. We use $E_0=M$ as a starting point and typically achieve the required tolerance in three or four iterations. If a solution to $E$ is not found within a default of fifty iterations, the procedure is restarted with $E_0=\pi$. If a solution is still not found, we instead use the numerical approximation described below to estimate $E$.

For highly-eccentric orbits ($e\sim1$) the number of iterations required to reach the required tolerance increases dramatically, especially as $M$ approaches 0 or $2\pi$, significantly reducing the speed of this solver (Fig.~\ref{fig:kepler}, left panel). This can be especially problematic for the parallel-tempered MCMC sampler described in Section~\ref{sec:sampler}, where the high-temperature walkers will explore the full range of eccentricities allowed by the prior distribution. To mitigate this slowdown, we instead use the numerical approximation for Kepler's equation described by \citet{Mikkola:1987aa} for orbits with $e\ge0.95$. Mikkola's algorithm invokes a cubic approximation to Kepler's equation and a numerically intensive (but single iteration) high-precision correction formula. 

We verified the accuracy of Mikkola's method by comparing the value of $E$ calculated for a range of $(M, e)$ combinations using Newton's method (with a tolerance of $10^{-18}$) and using Mikkola's method. We find a maximum absolute difference between the eccentric anomaly computed via both methods of $1.6\times10^{-15}$ (Fig.~\ref{fig:kepler}, middle panel), well below the nominal tolerance on the eccentric anomaly solver of $|E_n-E_{n-1}| < 10^{-9}$. While Mikkola's method is strictly more accurate than applying Newton's method with a tolerance of $10^{-9}$, it is much more computationally expensive, especially for low-eccentricity orbits where it is a factor of ten slower (Fig.~\ref{fig:kepler}, right panel). For the vast majority of applications for \orbitize the default tolerance of $10^{-9}$ corresponds to a position angle error of $3.6\times10^{-7}$\,deg for a face-on circular orbit, well below the precision of current astrometric measurements.

These algorithms are implemented in C for maximum computational efficiency. A Python version of this module is also included in \orbitize{}, and the package will revert to the Python version if the C-solver fails to compile on a user's machine. 

\subsubsection{Planetary System}
\label{sec:system}
The main component of the \texttt{orbitize.system} module is the \texttt{orbitize.system.System} class, which stores observational data, priors, and posteriors for a star-planet system. To initialize an instance of \texttt{orbitize.system.System}, a user inputs a data table, created using the \texttt{orbitize.read\_input} module, a total system mass, and a parallax. The \texttt{\_\_init\_\_()} method automatically initializes a list of \texttt{orbitize.priors.Prior} objects in a standard order, detailed in the online documentation. \texttt{orbitize.system.System} also has a \texttt{compute\_model()} method that takes in an array of potential orbital parameters and computes model predictions to be compared against the data. This is the method that calls the Keplerian orbit-solver (Section \ref{sec:kepler}). Note that this abstraction makes the sampler agnostic of physics; a user could use any model computation code as a drop-in replacement. Finally, this class has an attribute \texttt{results}, a list of \texttt{orbitize.results.Results} objects (Section \ref{sec:results}). In version \currentversion{}, \texttt{orbitize!} is limited to single-planet systems. Users can create \texttt{orbitize.system.System} objects for each planet in a multi-planet system, but orbit-fits using this framework currently do not take into account the dynamical effects of other planets on the orbit fit.

\subsubsection{Results}
\label{sec:results}
The \texttt{orbitize.results} module contains code for the \texttt{orbitize.results.Results} class. Each instance of this class represents a posterior (whether calculated using OFTI or MCMC). The \texttt{orbitize.sampler.Sampler} objects add their output to an instance of this class. This module also controls saving, loading, and visualizing results. 

%Then, a user can use methods in this class to save orbit samples to a file, to load saved orbit samples, or to visualize the orbit samples through a few plotting routines. Samples, along with some metadata, can be saved to a file in either \texttt{hdf5} (default and recommended) or \texttt{FITS} format. To retrieve these samples later, a user could initialize a \texttt{orbitize.results.Results} object and then use one of the object's methods to load the samples from the file into the object. 

% There are two plotting routines. The first plotting routine is \texttt{orbitize.results.Results.plot\_corner()} which uses \texttt{corner.py} \red{(cite)} to plot two-dimensional marginalized posteriors for each pair of parameters \red{(add a figure example? or show this in Section 3.2?)}. The second plotting routine is \texttt{orbitize.results.Results.plot\_orbits()} which shows one orbital period in two-dimensional space for a selected number of orbits in the sample set. \red{(add a figure example? or show this in Section 3.2?)}

\subsubsection{Samplers}
\label{sec:sampler}
\orbitize{} implements the OFTI algorithm (Sec \ref{sec:OFTI_algorithm}) in the \texttt{orbitize.sampler.OFTI} class and two MCMC algorithms (Sec \ref{sec:MCMC_algorithm}) in the \texttt{orbitize.sampler.MCMC} class.

The \texttt{run\_sampler()} method of the \texttt{orbitize.sampler.OFTI} class generates a posterior of permissible orbits. This method iteratively runs a series of methods that define the process of generating and statistically vetting potential orbits until a desired number of orbits are accepted. Accepted orbits are then added to an instance of the \texttt{orbitize.Results} class. This process is arbitrarily parallelizible, and users can easily set the number of CPU cores available for fits. 

% Once the \texttt{orbitize.sampler.OFTI} class is initialized, the \texttt{prepare\_samples()} method within it generates a number of sample orbits for the system. This method takes in a desired number of generated orbits and returns an array of randomly generated orbits whose semi-major axis and position angle of nodes have been scaled-and-rotated to be consistent with input observations. This array can then be passed through a second method, \texttt{orbitize.sampler.OFTI.reject()}, which requires an array of sample orbits as input. The \texttt{reject()} method computes the probability of each orbit based on its associated chi squared, and rejects orbits with likelihood probabilities less than a uniform randomly generated number. The output of this method is an array of accepted orbits, and a second array containing the corresponding log likelihoods for each orbit.

% An alternative to the process described above is a third method within \texttt{orbitize.sampler.OFTI} called \texttt{run\_sampler()}. This method runs the \texttt{prepare\_samples()} and \texttt{reject()} methods repeatedly until a desired number of orbits are accepted. To call \texttt{run\_sampler}, the user inputs the number of samples to be generated each time \texttt{prepare\_samples()} is initialized, along with the number of accepted orbits to be returned. The final output is an array the user-specified number of accepted orbits and their associated likelihoods. Additionally, the \texttt{orbitize.Results} class will be updated to include all accepted orbits.

\texttt{orbitize.sampler.MCMC} uses MCMC algorithms to generate a Markov Chain representing the posterior. Setting the attribute \texttt{num\_temps} to 1 invokes the Affine-Invariant sampler, and setting it to a number greater than 1 invokes the parallel tempered sampler. Users also choose the number of walkers, the number of threads, and whether they would like to fix certain parameters so that they are not sampled by the MCMC algorithm. A convenient API for inspecting chains to assess convergence is available, and is explained in the online tutorials and documentation.

% Within the \texttt{orbitize.sampler.MCMC} class, there are methods to compute the log likelihood of a proposed orbit and to fill in values of parameters fixed by the user and not sampled. The main method is \texttt{orbitize.sampler.MCMC.run\_sampler()}. This method will create a \texttt{emcee} or \texttt{ptemcee} sampler object and run for the number of steps determined by a user. A user can also set the number of steps to discard as ``burn-in`` and whether they want to thin the final samples. After burn-in and thinning, the current position, posterior distributions (i.e. the ``chain'') and likelihood for each step in the chain are saved to the \texttt{orbitize.sampler.MCMC} object. If the Parallel Tempering sampler is used, only the lowest temperature chain and associated likelihoods are saved. The chain and likelihoods are also used to create an \texttt{orbitize.results.Results} object, which is added to the \texttt{orbitize.sampler.MCMC} object as well.

\subsubsection{Driver}
Finally, the \texttt{orbitize.driver} module automates the creation of the data table and \texttt{orbitize.system.System} objects. It is a convenient short-cut for standard orbit-fits, and allows new users to begin using the code relatively quickly. While using this object makes it simple to run \texttt{orbitize!} in a standard way, users are encouraged to learn about the underlying API in order to learn how to, e.g. modify priors, customize plots, and set parameters specific to OFTI or MCMC. 

\subsection{Example Usage}
\label{sec:usage}

To illustrate the ease with which a user can run an orbit fit with \texttt{orbitize!}, we provide a minimal code example below. One of the strengths of \texttt{orbitize!} is its customizability, however, and this example is merely intended to show how easy it is to run a ``standard'' orbit fit, not to give a sense of the full scale of the code's capabilities. We encourage users to peruse the online tutorials\footnote{\href{http://orbitize.info/en/latest/tutorials.html}{http://orbitize.info/en/latest/tutorials.html}} for more in-depth examples of how to use and modify \texttt{orbitize!}.

\begin{figure*}
    \centerline{\includegraphics[width=1.3\textwidth]{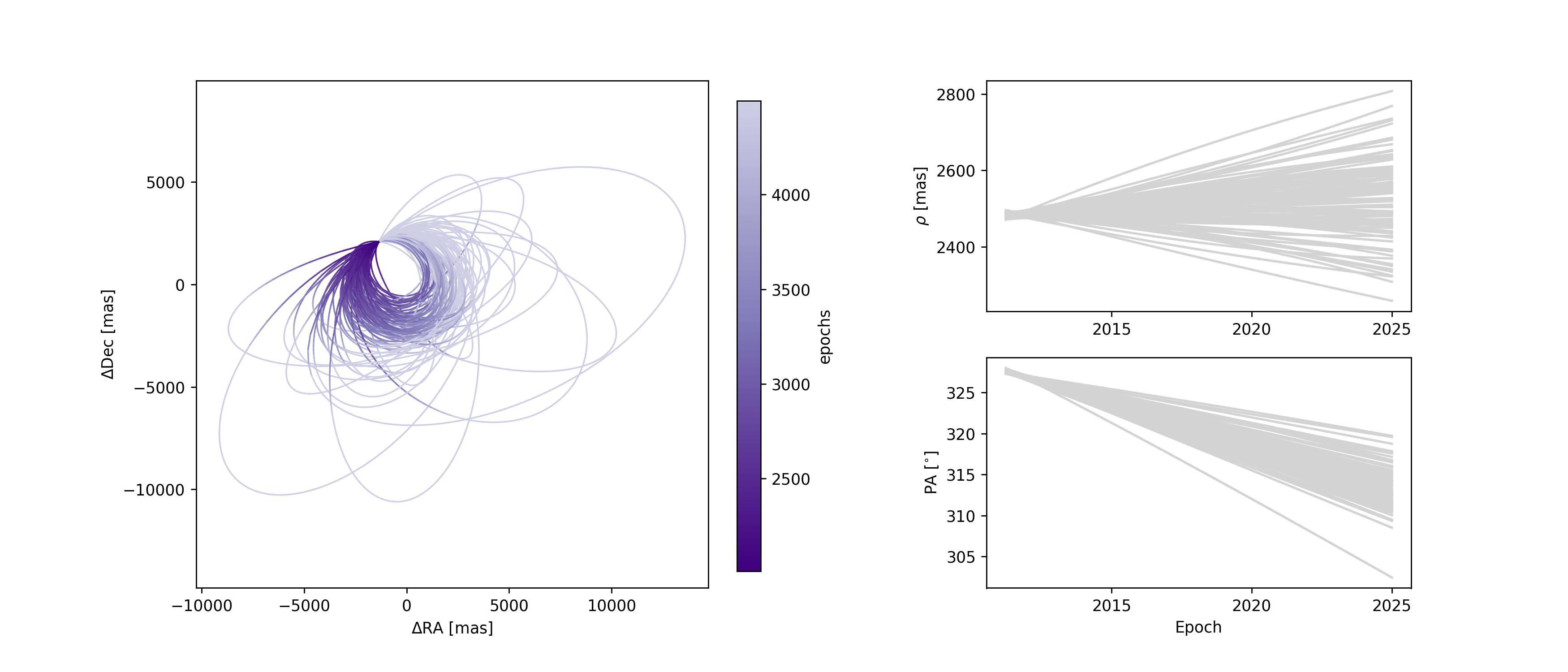}}
    \caption{Orbit figure produced from the code snippet.}
    \label{fig:orbit_fig}
\end{figure*}

\begin{figure*}
    \centerline{\includegraphics[width=1.1\textwidth]{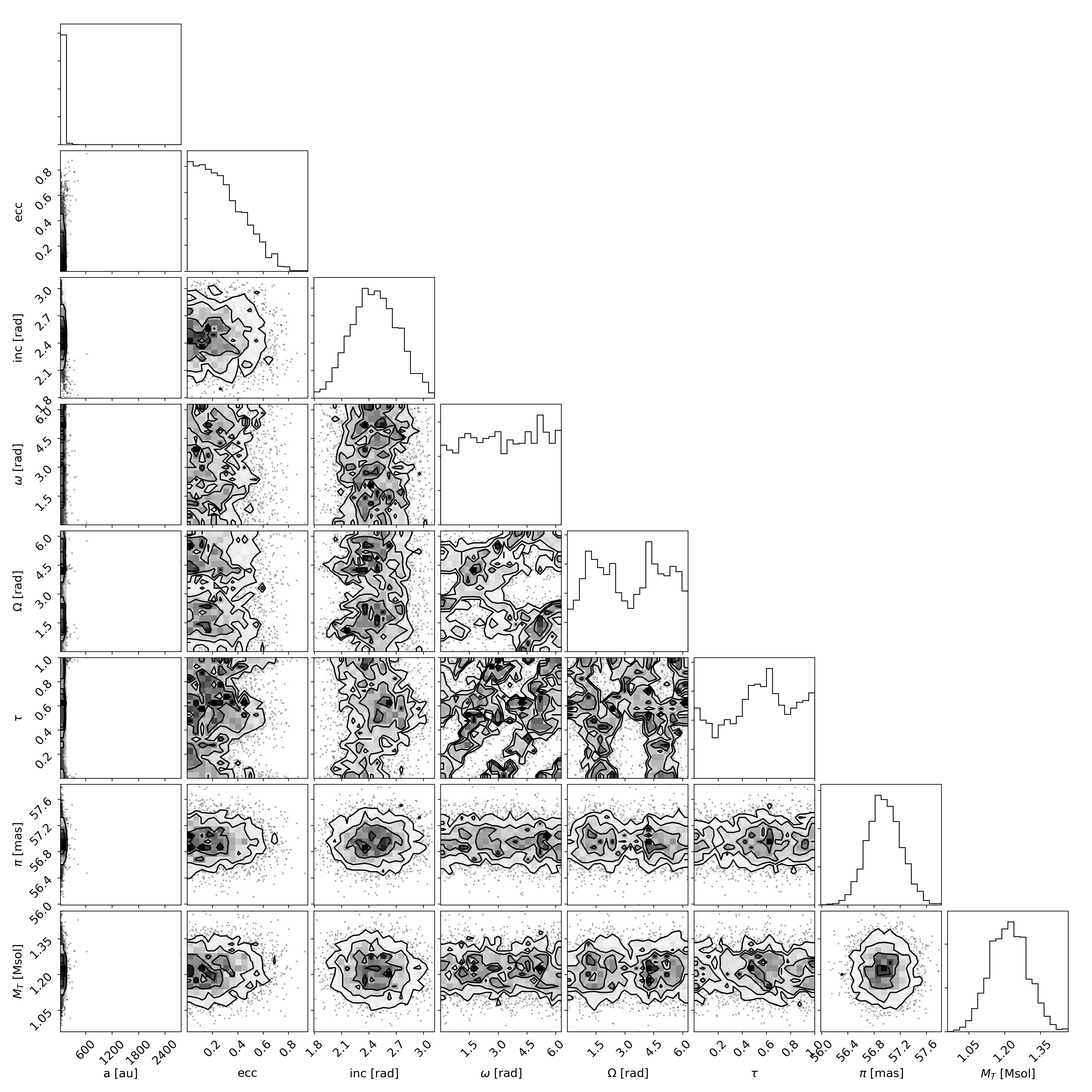}}
    \caption{Corner plot produced from the code snippet. This plot was produced using \texttt{corner} \citep{corner}.}
    \label{fig:corner_fig}
\end{figure*}

In the example below, an \texttt{orbitize.driver.Driver} object is initialized, and the OFTI algorithm runs until 10,000 orbits have been accepted. To use MCMC instead of the OFTI algorithm, the user just needs to switch out the keyword in line 6. Inline comments are provided to aid understanding. The two figures produced by this code snippet are shown in Figures \ref{fig:orbit_fig} and \ref{fig:corner_fig}. 

\begin{lstlisting}[language=Python]
from orbitize import driver
from orbitize import DATADIR

myDriver = driver.Driver(
    '{}/GJ504.csv'.format(DATADIR), # data file
    'OFTI',        # choose from: ['OFTI', 'MCMC']
    1,             # number of planets in system
    1.22,          # total system mass [M_sun]
    56.95,         # system parallax [mas]
    mass_err=0.08, # mass error [M_sun]
    plx_err=0.26   # parallax error [mas]
)
orbits = myDriver.sampler.run_sampler(10000)

# plot the results
myResults = myDriver.sampler.results
orbit_figure = myResults.plot_orbits(
# minimum MJD for colorbar (choose first data epoch)
    start_mjd=myDriver.sampler.epochs[0] 
)
corner_figure = myResults.plot_corner()
\end{lstlisting}

\section{Future Prospects}
\label{sec:future}
\subsection{Community Involvement Guidelines}
\label{sec:community}

In order to realize our goals of consolidating direct imaging orbit-fitting best practices in one place and continually adapting \orbitize{} to serve the direct imaging community, we require substantial community involvement. We strongly encourage anyone interested in using \orbitize{} to contribute code. We maintain a document with specific contributor guidelines\footnote{\href{https://github.com/sblunt/orbitize/blob/master/contributor\_guidelines.md}{https://github.com/sblunt/orbitize/blob/master/\\contributor\_guidelines.md}} on our GitHub page. 

In addition to directly contributing code, we encourage community members to request features, report bugs, and provide feedback through raising issues on GitHub. This is the most efficient way to reach the entire development team.

\subsection{Future Work}
\label{sec:todo}

\orbitize{} has a long and exciting list of planned updates. A version 2.0 release, intended to implement many of these, is planned for 2020. Key upgrades in version 2.0 will be enabling dynamical mass measurements by jointly fitting RV and stellar astrometry datasets, as well as fitting multi-planet systems. To summarize, the major upcoming features are:

\begin{enumerate}
    \item Jointly fitting radial velocities
    \item Incorporating independent radial velocity fits as priors
    \item Jointly fitting stellar astrometry
    \item Fitting multi-planet and hierarchical systems
    \item Fitting using other orbital element parametrizations (e.g. $\sqrt{e}\cos{\omega}$ and $\sqrt{e}\sin{\omega}$ rather than $e$ and $\omega$)
    \item Incorporating an N-body integrator to replace the standard Kepler solver where three-body interactions are nonnegligible
    \item Accounting for instrumental calibration systematics
    \item Adding a Hamiltonian MCMC algorithm backend
    \item Incorporating observation-driven priors (e.g. \citealt{ONeil:2019aa})
\end{enumerate}

\section{Conclusion}
\label{sec:concl}

In this paper, we have presented \orbitize{}, an open-source Python package for fitting the orbits of directly-imaged planets. \orbitize{} uses OFTI and MCMC, two efficient, Bayesian methods for computing posteriors. We aim to encourage community contributions, remove barriers to becoming an expert in orbit-fitting, and provide an open-source development environment in order to meet the orbit-fitting needs of the high-contrast exoplanet imaging community.

\acknowledgements{

The authors thank those at academic and telescope facilities whose labor maintains spaces for scientific inquiry, particularly those whose communities are excluded from the academic system.

This package was born and developed at the winter 2018 and 2019 AAS Hack Days, and the authors wish to thank the organizers of these events. The authors would also like to thank Dillon Dong, Jasmine Garani, Melisa Tallis, and Daniel Yahalomi for their time and initial work on \texttt{orbitize!}, and Junellie Gonzalez, Kelly Kosmo O'Neil, Ryan Rubenzahl, and Jean-Baptiste Ruffio for participating in our hackathons and for their anticipated future contributions to \orbitize{} Finally, we thank the anonymous individual who unknowingly named \texttt{orbitize!} at the 2018 AAS Hack Day. 

Yes, you need to use the exclamation point!

}

\bibliographystyle{aasjournal}
\bibliography{biblio,supp_refs}

\begin{thebibliography}{}
\expandafter\ifx\csname natexlab\endcsname\relax\def\natexlab#1{#1}\fi

\bibitem[{{Bate} {et~al.}(2010){Bate}, {Lodato}, \& {Pringle}}]{Bate:2010aa}
{Bate}, M.~R., {Lodato}, G., \& {Pringle}, J.~E. 2010, \mnras, 401, 1505

\bibitem[{{Binnendijk}(1960)}]{Binnendijk:1960aa}
{Binnendijk}, L. 1960, {Properties of double stars; a survey of parallaxes and
  orbits.}

\bibitem[{{Blunt} {et~al.}(2017){Blunt}, {Nielsen}, {De Rosa}, {Konopacky},
  {Ryan}, {Wang}, {Pueyo}, {Rameau}, {Marois}, {Marchis}, {Macintosh},
  {Graham}, {Duch{\^e}ne}, \& {Schneider}}]{Blunt:2017aa}
{Blunt}, S., {Nielsen}, E.~L., {De Rosa}, R.~J., {et~al.} 2017, \aj, 153, 229

\bibitem[{{Brandt} {et~al.}(2018){Brandt}, {Dupuy}, \&
  {Bowler}}]{Brandt:2018aa}
{Brandt}, T.~D., {Dupuy}, T.~J., \& {Bowler}, B.~P. 2018, arXiv e-prints,
  arXiv:1811.07285

\bibitem[{{Dawson} {et~al.}(2011){Dawson}, {Murray-Clay}, \&
  {Fabrycky}}]{Dawson:2011aa}
{Dawson}, R.~I., {Murray-Clay}, R.~A., \& {Fabrycky}, D.~C. 2011, \apjl, 743,
  L17

\bibitem[{{Eastman} {et~al.}(2013){Eastman}, {Gaudi}, \&
  {Agol}}]{Eastman:2013aa}
{Eastman}, J., {Gaudi}, B.~S., \& {Agol}, E. 2013, \pasp, 125, 83

\bibitem[{Foreman-Mackey(2016)}]{corner}
Foreman-Mackey, D. 2016, The Journal of Open Source Software, 24,
  doi:10.21105/joss.00024

\bibitem[{{Foreman-Mackey} {et~al.}(2013){Foreman-Mackey}, {Hogg}, {Lang}, \&
  {Goodman}}]{Foreman-Mackey:2013aa}
{Foreman-Mackey}, D., {Hogg}, D.~W., {Lang}, D., \& {Goodman}, J. 2013, \pasp,
  125, 306

\bibitem[{{Fulton} {et~al.}(2018){Fulton}, {Petigura}, {Blunt}, \&
  {Sinukoff}}]{Fulton:2018aa}
{Fulton}, B.~J., {Petigura}, E.~A., {Blunt}, S., \& {Sinukoff}, E. 2018, \pasp,
  130, 044504

\bibitem[{{Ginski} {et~al.}(2013){Ginski}, {Neuh{\"a}user}, {Mugrauer},
  {Schmidt}, \& {Adam}}]{Ginski:2013aa}
{Ginski}, C., {Neuh{\"a}user}, R., {Mugrauer}, M., {Schmidt}, T.~O.~B., \&
  {Adam}, C. 2013, \mnras, 434, 671

\bibitem[{{Gravity Collaboration} {et~al.}(2019){Gravity Collaboration},
  {Lacour}, {Nowak}, {Wang}, {Pfuhl}, {Eisenhauer}, {Abuter}, {Amorim},
  {Anugu}, {Benisty}, {Berger}, {Beust}, {Blind}, {Bonnefoy}, {Bonnet},
  {Bourget}, {Brandner}, {Buron}, {Collin}, {Charnay}, {Chapron}, {Cl{\'e}net},
  {Coud{\'e} Du Foresto}, {de Zeeuw}, {Deen}, {Dembet}, {Dexter}, {Duvert},
  {Eckart}, {F{\"o}rster Schreiber}, {F{\'e}dou}, {Garcia}, {Garcia Lopez},
  {Gao}, {Gendron}, {Genzel}, {Gillessen}, {Gordo}, {Greenbaum}, {Habibi},
  {Haubois}, {Hau{\ss}mann}, {Henning}, {Hippler}, {Horrobin}, {Hubert},
  {Jimenez Rosales}, {Jocou}, {Kendrew}, {Kervella}, {Kolb}, {Lagrange},
  {Lapeyr{\`e}re}, {Le Bouquin}, {L{\'e}na}, {Lippa}, {Lenzen}, {Maire},
  {Molli{\`e}re}, {Ott}, {Paumard}, {Perraut}, {Perrin}, {Pueyo}, {Rabien},
  {Ram{\'\i}rez}, {Rau}, {Rodr{\'\i}guez-Coira}, {Rousset}, {Sanchez-Bermudez},
  {Scheithauer}, {Schuhler}, {Straub}, {Straubmeier}, {Sturm}, {Tacconi},
  {Vincent}, {van Dishoeck}, {von Fellenberg}, {Wank}, {Waisberg}, {Widmann},
  {Wieprecht}, {Wiest}, {Wiezorrek}, {Woillez}, {Yazici}, {Ziegler}, \&
  {Zins}}]{Gravity-Collaboration:2019aa}
{Gravity Collaboration}, {Lacour}, S., {Nowak}, M., {et~al.} 2019, \aap, 623,
  L11

\bibitem[{{Green}(1985)}]{Green:1985aa}
{Green}, R.~M. 1985, {Spherical Astronomy}

\bibitem[{{Lagrange} {et~al.}(2010){Lagrange}, {Bonnefoy}, {Chauvin}, {Apai},
  {Ehrenreich}, {Boccaletti}, {Gratadour}, {Rouan}, {Mouillet}, {Lacour}, \&
  {Kasper}}]{Lagrange:2010aa}
{Lagrange}, A.~M., {Bonnefoy}, M., {Chauvin}, G., {et~al.} 2010, Science, 329,
  57

\bibitem[{{Maire} {et~al.}(2019){Maire}, {Rodet}, {Cantalloube}, {Galicher},
  {Brandner}, {Messina}, {Lazzoni}, {Mesa}, {Melnick}, {Carson}, {Samland},
  {Biller}, {Boccaletti}, {Wahhaj}, {Beust}, {Bonnefoy}, {Chauvin}, {Desidera},
  {Langlois}, {Henning}, {Janson}, {Olofsson}, {Rouan}, {M{\'e}nard},
  {Lagrange}, {Gratton}, {Vigan}, {Meyer}, {Cheetham}, {Beuzit}, {Dohlen},
  {Avenhaus}, {Bonavita}, {Claudi}, {Cudel}, {Daemgen}, {D'Orazi}, {Fontanive},
  {Hagelberg}, {Le Coroller}, {Perrot}, {Rickman}, {Schmidt}, {Sissa}, {Udry},
  {Zurlo}, {Abe}, {Orign{\'e}}, {Rigal}, {Rousset}, {Roux}, \&
  {Weber}}]{Maire:2019aa}
{Maire}, A.~L., {Rodet}, L., {Cantalloube}, F., {et~al.} 2019, \aap, 624, A118

\bibitem[{{Mikkola}(1987)}]{Mikkola:1987aa}
{Mikkola}, S. 1987, Celestial Mechanics, 40, 329

\bibitem[{{Mugrauer} {et~al.}(2012){Mugrauer}, {R{\"o}ll}, {Ginski}, {Vogt},
  {Neuh{\"a}user}, \& {Schmidt}}]{Mugrauer:2012aa}
{Mugrauer}, M., {R{\"o}ll}, T., {Ginski}, C., {et~al.} 2012, \mnras, 424, 1714

\bibitem[{{O'Neil} {et~al.}(2019){O'Neil}, {Martinez}, {Hees}, {Ghez}, {Do},
  {Witzel}, {Konopacky}, {Becklin}, {Chu}, {Lu}, {Matthews}, \&
  {Sakai}}]{ONeil:2019aa}
{O'Neil}, K.~K., {Martinez}, G.~D., {Hees}, A., {et~al.} 2019, \aj, 158, 4

\bibitem[{{Scharf} \& {Menou}(2009)}]{Scharf:2009aa}
{Scharf}, C., \& {Menou}, K. 2009, \apjl, 693, L113

\bibitem[{{Seager}(2010)}]{Seager:2010aa}
{Seager}, S. 2010, {Exoplanets}

\bibitem[{{Sozzetti} {et~al.}(1998){Sozzetti}, {Spagna}, \&
  {Lattanzi}}]{Sozzetti:1998aa}
{Sozzetti}, A., {Spagna}, A., \& {Lattanzi}, M.~G. 1998, Earth Moon and
  Planets, 81, 103

\bibitem[{{Th{\'e}bault} \& {Beust}(2001)}]{Thebault:2001aa}
{Th{\'e}bault}, P., \& {Beust}, H. 2001, \aap, 376, 621

\bibitem[{{Vigan} {et~al.}(2016){Vigan}, {Bonnefoy}, {Ginski}, {Beust},
  {Galicher}, {Janson}, {Baudino}, {Buenzli}, {Hagelberg}, {D'Orazi},
  {Desidera}, {Maire}, {Gratton}, {Sauvage}, {Chauvin}, {Thalmann}, {Malo},
  {Salter}, {Zurlo}, {Antichi}, {Baruffolo}, {Baudoz}, {Blanchard},
  {Boccaletti}, {Beuzit}, {Carle}, {Claudi}, {Costille}, {Delboulb{\'e}},
  {Dohlen}, {Dominik}, {Feldt}, {Fusco}, {Gluck}, {Girard}, {Giro}, {Gry},
  {Henning}, {Hubin}, {Hugot}, {Jaquet}, {Kasper}, {Lagrange}, {Langlois}, {Le
  Mignant}, {Llored}, {Madec}, {Martinez}, {Mawet}, {Mesa}, {Milli},
  {Mouillet}, {Moulin}, {Moutou}, {Orign{\'e}}, {Pavlov}, {Perret}, {Petit},
  {Pragt}, {Puget}, {Rabou}, {Rochat}, {Roelfsema}, {Salasnich}, {Schmid},
  {Sevin}, {Siebenmorgen}, {Smette}, {Stadler}, {Suarez}, {Turatto}, {Udry},
  {Vakili}, {Wahhaj}, {Weber}, \& {Wildi}}]{Vigan:2016aa}
{Vigan}, A., {Bonnefoy}, M., {Ginski}, C., {et~al.} 2016, \aap, 587, A55

\bibitem[{{Vousden} {et~al.}(2016){Vousden}, {Farr}, \&
  {Mandel}}]{Vousden:2016aa}
{Vousden}, W.~D., {Farr}, W.~M., \& {Mandel}, I. 2016, \mnras, 455, 1919

\bibitem[{{Wagner} {et~al.}(2016){Wagner}, {Apai}, {Kasper}, {Kratter},
  {McClure}, {Robberto}, \& {Beuzit}}]{Wagner:2016aa}
{Wagner}, K., {Apai}, D., {Kasper}, M., {et~al.} 2016, Science, 353, 673

\bibitem[{{Wagner} {et~al.}(2018){Wagner}, {Dong}, {Sheehan}, {Apai}, {Kasper},
  {McClure}, {Morzinski}, {Close}, {Males}, {Hinz}, {Quanz}, \&
  {Fung}}]{Wagner:2018aa}
{Wagner}, K., {Dong}, R., {Sheehan}, P., {et~al.} 2018, \apj, 854, 130

\bibitem[{{Wertz} {et~al.}(2017){Wertz}, {Absil}, {G{\'o}mez Gonz{\'a}lez},
  {Milli}, {Girard}, {Mawet}, \& {Pueyo}}]{Wertz:2017aa}
{Wertz}, O., {Absil}, O., {G{\'o}mez Gonz{\'a}lez}, C.~A., {et~al.} 2017, \aap,
  598, A83

\bibitem[{{Williams} \& {Pollard}(2002)}]{Williams:2002aa}
{Williams}, D.~M., \& {Pollard}, D. 2002, International Journal of
  Astrobiology, 1, 61

\bibitem[{{Yu} \& {Tremaine}(2001)}]{Yu:2001aa}
{Yu}, Q., \& {Tremaine}, S. 2001, \aj, 121, 1736

\bibitem[{{Zieba} {et~al.}(2019){Zieba}, {Zwintz}, {Kenworthy}, \&
  {Kennedy}}]{Zieba:2019aa}
{Zieba}, S., {Zwintz}, K., {Kenworthy}, M.~A., \& {Kennedy}, G.~M. 2019, \aap,
  625, L13

\end{thebibliography}

\end{document}